# Spin as Dynamic Variable

## or

## Why Parity is Broken

*G. N. Golub'*
golubgn@meta.ua

There suggested a modification of the Dirac electron theory, eliminating its mathematical incompleteness. The modified Dirac electron, called dual, is described by two waves, one of which is the Dirac wave and the second dynamically restores the symmetry unused in the Dirac theory of the Dirac equation. There has been produced a set of the conserved currents which are generated by equations of the dual electron. It is shown that, although, it consists of the vector, axial and left currents, nevertheless, it *has mirror symmetry*. It is also shown that if the observed electron follows the dual electron equations it would *violate the mirror symmetry*. The results of the paper for the first time explain the ways and the reasons of mirror symmetry being created and broken in the nature.

## 1. Introdaction

Dirac wave equation [1 - 4]

$$(i\gamma \cdot \partial - m)\psi(x) = 0 \qquad (1)$$

has been intended for the electron description with the spin, co-operating electromagnetic forces. A vector conserved current

$$J^\mu(x) = \bar{\psi}(x)\gamma^\mu\psi(x), \qquad (2)$$

generated by this equation, at inclusion of electromagnetic interaction becomes an electromagnetic current of an electron. On the basis of Dirac equation and Maxwell equations within quantum electrodynamics almost complete description of electromagnetic properties of an observable electron is reached. But for the description of weak interactions in which the observable electron participates, it is necessary to use the currents which conservation is not generated by Dirac equation (1). In Standard model of electroweak interactions[5 - 8], in which the description of weak interactions of an observable electron is reached, the spatial structure of its weak currents should be taken from experience. As weak interactions break mirror symmetry, it is represented impossible, that based on specularly symmetric equation (1) the Dirac theory of an electron could generate weak interacting currents of an electron.

But what is represented impossible from one point of view, can appear possible from some other point of view. To come to such other point of view it is necessary to distinguish, first, Dirac wave equation and the Dirac's theory of an electron based on it. The last includes, besides the wave equation, two more spinor equations, algebra of the Dirac $\gamma$-matrixes and the geometrical theory of Dirac bispinor generated by them. Wave equation (1) really it is not capable to generate the left remaining current. But in the Dirac theory of an electron there can be enough place and possibilities to generate not only vector conserved currents. Secondly, physical interpretation of the Dirac theory – "a picture of a Dirac electron" – extremely strongly limits the use of the mathematical formalism, actually hiding its true possibilities. The picture of a Dirac electron "is ground" for the description *of an electromagnetic* electron and it appears incompatible with attempt to describe *an* *electroweak* electron, using a mathematical formalism of the Dirac theory. If we set such purpose, existing interpretation should be modified. Originally to modification should be subject, on - to a being, only one, but key concept of the Dirac theory – electron spin. As a result of modification of spin should turn from the nomenclature size  only indirectly influencing dynamic processes, into the real dynamic variable, same, as an electron impulse. The initial  mathematical structure of the Dirac theory remains thus without changes –only *sense* and *roles* of the equations will change. As a result it will become clear  that violation of parity  which we observe in the nature, contains – in the hidden look - in the Dirac theory of an electron. In it  some "secret" side  of the equation of Dirak  also consists. Sense of the real work is in its disclosure.

## 2 . The Dirac spinor equation

Unlike other fundamental equations of physics using language of the analysis  and geometry, Dirac equation uses analysis and algebra language. . Because of this, in the Dirac's theory appear *two* of the Dirac equations: one is differential or wave, the other – algebraic or spinor. The latter is usually called "the Dirac equation in the impulse representation". Having substituted in the wave equation (1) the decision for a plane wave with both signs of frequencies

$$\psi_{\pm p}(x) = e^{\mp i p \cdot x} U_{\pm}(p), \qquad (3)$$

let's receive the Dirac spinor equation:

$$(\gamma \cdot p \mp m) U_{\pm}(p) = 0. \qquad (4)$$

Here $U_{\pm}(p)$ - Dirac bispinor, $p^{\mu}$ and $m-$ impulse and mass of an electron. The interfaced equation looks like:

$$\bar{U}_{\pm}(p)(\gamma \cdot p \mp m) = 0. \qquad (5)$$

Dirac  spinor current $j^{\mu}$ is defined by expression

$$j^{\mu} = \bar{U}_{\pm}(p') \gamma^{\mu} U_{\pm}(p) \qquad (6)$$

also satisfies to a condition received by a standard method from (4) and (5):

$$(p - p') \cdot j = 0. \qquad (7)$$

Also  from  Dirac spinor  equation  and  the  interfaced equation it is possible to receive ratio

$$\bar{U}_{\pm}(p) \gamma^{\mu} U_{\pm}(p) = \pm \frac{p^{\mu}}{m} \bar{U}_{\pm}(p) U_{\pm}(p), \qquad (8)$$

from which we will receive normalization of bispinors

$$\bar{U}_{\pm}(p) U_{\pm}(p) = \pm 1 \qquad (9)$$

and expression of a free spinor current in four-dimensional speed of an electron $v^{\mu} = p^{\mu} m^{-1}$

$$v^{\mu} = \bar{U}_{\pm}(p) \gamma^{\mu} U_{\pm}(p). \qquad (10)$$

Above we followed a usual and most common way of creation of the Dirac theory[4, 9 – 11], consisting  that Dirac spinor equation is deduced  from the wave equation (1). Meets as well an alternative way  at which start with the spinor equations and then pass to the wave[3, 12]. In the real work we will follow an alternative way of creation of the theory, though  have begun with the usual. So, we will accept  that the initial equation of the Dirac theory is Dirac spinor equation (4). To deduce from it Dirac wave equation, we should, operating formally, to make

$$p^\mu \to i\partial^\mu, U_\pm(p) \to \psi_{\pm p}(x) \qquad (11)$$

and to receive as a result of plane wave Dirac equation. Then, using a superposition principle [3], to receive the wave equation (1). But for us it is important that at the carried-out replacement a spinor current (6) passes in plane wave current

$$j^\mu \to J^\mu_{pl}(x) = \bar\psi_{\pm p'}(x)\gamma^\mu\psi_{\pm p}(x) \qquad (12)$$

and equality (7) passes to the law of preservation of plane wave current

$$\partial \cdot J_{pl}(x) = 0. \qquad (13)$$

Therefore we will call equality (7) *law of conservation of a spinor current*.

Now we approach to central point of our consideration of the Dirac theory. It consists that Dirac spinor equation short defines a bispinor $U_\pm(p)$, leaving uncertain a half its component. In a picture of a Dirac electron it has a natural explanation: an electron has a spin which behaviour is not defined neither wave, nor spinor by Dirac equations. What does it mean in sense of use of degrees of freedom available for a Dirac electron? It means, simply saying that in dynamic sense the Dirac electron is half empty. In itself it would not cause any questions if the observable electron co-operated only electromagnrtically and as it would well be described by a Dirac electron as it occurs now. But the observable electron, except electromagnetic, co-operates with weak forces, for what the existing Dirac electron is not capable. (Here, to a possible question of the reader: "Doesn't the Standard model of electroweak interactions describes all interactions of a Dirac electron, both electromagnetic, and weak?" it is necessary to answer: "No, it does not describe. The Standard model is a *modification* of the Dirac theory, it describes the modified Dirac electron and it possesses all interactions of an observable electron.") It is comparable now with dynamic incompleteness of a Dirac electron and dynamic "overpopulation" – in comparison with Dirac – of an observable electron. There is a question:"Are they connected with each other, i.e. is incompleteness of a Dirac electron the reason of weak forcesat it?" It is possible that it is so, and then it is necessary to try *to complete* the Dirac theory of an electromagnetic electron to the complete dynamic theory in hope to receive *an electroweak* Dirac electron. Another part of this work is devoted to attempt of implementation of this hope.

The first that it is necessary to make for this purpose – to find the second spinor equation which together with Dirac spinor equation will make full system of the equations for a Dirac bispinor.

## 3. The equation of own polarisation

It, of course, should be the equation of spinor polarisation. Usually [4] bispinor polarisation is described by a four-dimensional *vector of polarisation* which, by definition, is equal in system of rest of a bispinor to a three-dimensional vector of spin to which in this system the tensor of a bispinor spin, in turn, is reduced. Also it is usual to call a vector of polarisation *a spin vector*. Though it is not absolutely correct, as spin as the angular moment is a tensor, but it is convenient and we will use also this name. The reason of its common use is that within a picture of a Dirac electron it is impossible to establish true sense of a vector of polarisation. This sense makes a theoretical riddle of the Dirac theory, to solve which it is possible only by its modification.

Let's call *a vector of own polarisation* of a bispinor $U_\pm(p,s)$ vector

$$s^\mu = \bar{U}_\pm(p,s)\gamma_A^\mu U_\pm(p,s). \tag{14}$$

Here $\gamma_A^\mu = \gamma^\mu \cdot \gamma_5$, $\gamma_5 = i\gamma^0\gamma^1\gamma^2\gamma^3$. From Dirac spinor equation follows the equity

$$\bar{U}_\pm(p,s)\gamma_5 U_\pm(p,s) = 0. \tag{15}$$

Using it after multiplication of the equation of Dirac at the left on $\bar{U}_\pm(p,s)\gamma_5$, we will receive

$$s \cdot p = 0. \tag{16}$$

From here comes

$$[\gamma \cdot p, \gamma_A \cdot s] = 0, \tag{17}$$

and from consideration $s^\mu$ in system of rest follows $s \cdot s = -1$. Therefore, it should be carried out the equation

$$\gamma_A \cdot s U_\pm(p,s) = \lambda_\pm U_\pm(p,s).$$

By multiplying it at the left by $\bar{U}_\pm(p,s)$, let's receive $s^2 = \pm \lambda_\pm$ and, it means, $\lambda_\pm = \mp 1$. Thus, equation of own polarisation of a Dirac bispinor looks like:

$$(\gamma_A \cdot s \pm 1)U_\pm(p,s) = 0, \tag{18}$$

and the interfaced equation – view

$$\bar{U}_\pm(p,s)(\gamma_A \cdot s \pm 1) = 0.$$

## 4. The equations of a dual bispinor

Let's write out the received full system of the equations with which should satisfy a Dirac bispinor

$$\begin{cases}(\gamma \cdot p \mp m)U_\pm(p,s) = 0 \\ (\gamma_A \cdot s \pm 1)U_\pm(p,s) = 0\end{cases}. \tag{19}$$

These are the last equations in our work, belonging to the existing Dirac theory, having a habitual appearance and intepretiruyemy within a picture of a Dirac electron. We will begin modification of the Dirac theory with release of this system from influence of a picture of a Dirac electron. For this purpose, first, we will divide the first equation on $m$. Secondly, we will get rid of concept of "a frequency sign", having replaced with this concept with *a bispinor sign*. Dirac bispinor, thus, can be *positive* and *negative*. Thirdly, we will enter designation $\gamma_V^\mu = \gamma^\mu$. Fourthly, we will replace designation of a vector of polarisation: $a^\mu = s^\mu$. As a result we will receive the following system of equations:

$$\begin{cases}(\gamma_V \cdot v \mp 1)U_\pm(v,a) = 0 \\ (\gamma_A \cdot a \pm 1)U_\pm(v,a) = 0\end{cases}. \tag{20}$$

To these equations it is necessary to add received before definition $v^\mu$ and $a^\mu$:

$$\begin{cases}v^\mu = \bar{U}_\pm(v,a)\gamma_V^\mu U_\pm(v,a) \\ a^\mu = \bar{U}_\pm(v,a)\gamma_A^\mu U_\pm(v,a)\end{cases}. \tag{21}$$

Systems of the equations (20) and (21), as it is obvious, possess symmetry of the top and bottom equations. Let's call this symmetry *dual*. In the existing Dirac theory this symmetry is ignored and collapses, apparently already from system (19). The purpose of our modification is to keep this symmetry and to take it as a principle definition of all mathematical and physical concepts of the modified theory. The source of dual symmetry – Dirac algebra, in it dual symmetry is expressed by symmetry of forming systems:

$$\begin{cases} \{\gamma_V^\mu, \gamma_V^\nu\} = 2g^{\mu\nu} \\ \{\gamma_A^\mu, \gamma_A^\nu\} = -2g^{\mu\nu} \end{cases}. \qquad (22)$$

Let's begin necessary definitions with system (21). According to the Dirac theory $v^\mu$ - four-dimensional speed of an electron, $a^\mu$ is its polarisation. According to dual symmetry so cannot be – or both of them are speeds, or they are polarisation. Actually, apparently, both of them and that, and another – in relation to a bispinor they are polarisation, in relation to an electron – speeds. For their distinction we will enter at first the general definition for vectors in Minkowsky spaces of various physical dimensions (impulses, polyarizatsiya, etc.), except the most co-ordinate space – time Minkowsky. Let's call vector $b^\mu$ *external*, if $b \cdot b > 0$. Let's call vector $d^\mu$ *internal*, if $d \cdot d < 0$. At last, let's call vector $c^\mu$ *boundary*, if $c \cdot c = 0$. These definitions replace terms time-like, spacelike and lightlike, respectively, expressing domination of habitual space – time. Now we will call $v^\mu$ *in the external speed* of an electron and *external polarisation* of a bispinor $U_\pm(v,a)$, and, respectively, $a^\mu$ - *in the internal speed* of an electron and *internal polarisation* of a bispinor.

Let's pass to system (20). Having increased its equations on $m$, let's receive the following system of equations:

$$\begin{cases} (\gamma_V \cdot p \mp m) W_\pm(p,q) = 0 \\ (\gamma_A \cdot q \pm m) W_\pm(p,q) = 0 \end{cases}. \qquad (23)$$

Here dual symmetry compels us to accept the following definitions. Vector $p^\mu = mv^\mu$ let's call *an external impulse* of an electron, a vector $q^\mu = ma^\mu$ let's call *an internal impulse* of an electron, a bispinor $W_\pm(p,q)$ let's call *dual*. System (23) there is a system of the equations of a dual bispinor. Let's write out separately the equations for positive and negative dual bispinor:

$$\begin{cases} (\gamma_V \cdot p - m) W_+(p,q) = 0 \\ (\gamma_A \cdot q + m) W_+(p,q) = 0 \end{cases} \begin{cases} (\gamma_V \cdot p + m) W_-(p,q) = 0 \\ (\gamma_A \cdot q - m) W_-(p,q) = 0 \end{cases}. \qquad (24)$$

Apparently from these equations, for transition from positive bispinor to negative it is necessary to change a weight sign. Therefore further we will study only positive bispinor, having entered designation $W = W_+(p,q)$. As it will be visible from further, all received results of work are transferred from positive bispinor on negative without changes.

Let's write finally initial equations for a positive bispinor and its interfaced:

$$\begin{cases} \bar{W}(\gamma_V \cdot p - m) = 0 \\ \bar{W}(\gamma_A \cdot q + m) = 0 \end{cases} \begin{cases} (\gamma_V \cdot p - m) W = 0 \\ (\gamma_A \cdot q + m) W = 0 \end{cases}. \qquad (25)$$

## 5 . Conserved currents of a dual bispinor

Let's compare number of degrees of freedom of Dirac and dual bispinor. Their geometrical number is identical and equal to five: three external plus three internal minus communication (16). But at a dual bispinor all are dynamic five, at Dirac – only three. As a result of a Dirac bispinor has one remaining current, dual as we will see, – six.

We now define the meaning of a *dual spinor current*. This current depends on two dual bispinors - initial $W$ and final $\bar{W}'$, which is the solution of the conjugated system, - and defined by:

$$j_M^\mu = \bar{W}' \gamma_M^\mu W. \tag{26}$$

Here, the index $M$ takes values $\{V, A, R, L\}$, and also there introduced the notations: $\gamma_R^\mu = \tfrac{1}{2}(\gamma_V^\mu + \gamma_A^\mu), \gamma_L^\mu = \tfrac{1}{2}(\gamma_V^\mu - \gamma_A^\mu)$. Thus, one pair of the dual bispinors generates four different currents differentiated by the space-time structure, only two of which are linearly independent. The essential difference of the Dirac spinor current from the dual spinor current depends on the first two impulses - $p$ and $p'$, and the second depends on four - $p, p', q, q'$. We now define the *conservation laws* for the dual spinor currents. As an example, we present two conservation laws, received by the standard methods from the system (25):

$$\begin{cases}(p-p')\cdot j_V = 0 \\ (q-q')\cdot j_A = 0\end{cases}. \tag{27}$$

The difference between the initial and final impulses which depends on the spinor current will be defined *the current impact*. Then, the first law of conservation out of (27), which is similar to a single conservation law of the Dirac spinor current, claims that the vector current is orthogonal to its external impact. Similarly, the second law states that the axial current is orthogonal to its internal impact. The law of conservation depends on the orthogonal nature of the spinor current and its impact. We now define two scalar values for dual bispinors – a true scalar $S = \bar{W}'W$ and pseudoscalar $P = \bar{W}'\gamma_5 W$, which, as well as the currents, depends on four impulses. If the current is not orthogonal to its impacts, and their production is proportional to either $S$, or $P$, in this case we define this equation as a *non-conservation law*. Although, the spinor currents have no apparent dual structure, their laws of conservation / non-conservation have a clear dual structure defined by the current impact. It is convenient to represent the dual structure of conservation/ non-conservation laws in the form of 2 × 2 matrix, with the external and internal laws located on its diagonal, and its antidiagonal having the raising laws, the impact of $q - p'$, and the lowering ones, with exposure to $p - q'$.

We proceed to finding all of the conservation laws for dual spinor currents. If we assume the equation system (25) independent , i.e., written for two different Dirac bispinors, these equations will generate two independent conserved Dirac currents - external and internal, and a set of conservation laws (27) for such a system will be complete. Though, the set of diagonal laws (27) may not be complete for the dual bispinor. The dual bispinor equations can generate such relations between currents and impact that under certain conditions some non-conservation laws can become conservation laws. We call such conservation laws *conditional*, unlike *unconditional* laws (27). Our problem now touches the question whether the dual bispinor generates conditional conservation laws and what their structure is.

To solve this question, we write all the products of the impulses and currents resulting from the system (25):

$$\begin{cases}p\cdot j_V = p'\cdot j_V = mS \\ p\cdot j_A = -p'\cdot j_A = -mP\end{cases} \begin{cases}q\cdot j_V = -q'\cdot j_V = mP \\ q\cdot j_A = q'\cdot j_A = -mS\end{cases}. \tag{28}$$

Using these relations, we find that all the laws of conservation / non-conservation are generated by the system (2). We write the result in the form of matrices with the dual structure where each cell contains two equations. There represented two of such matrices; one we will collect all the scalar equations or contains $S$ and the second - all the pseudoscalar or contains $P$. Besides, to ensure symmetry, including the mirror one, we add similar equations to these matrices, which have the sums of impulses instead of the impacts. The "scalar" matrix of such equations presented as follows:

$$\begin{pmatrix} \begin{cases} (p-p')\cdot j_V = 0 \\ (p+p')\cdot j_V = 2mS \end{cases} & \begin{cases} (q-p')\cdot j_R = -mS \\ (q+p')\cdot j_L = mS \end{cases} \\ \begin{cases} (p-q')\cdot j_R = mS \\ (p+q')\cdot j_L = mS \end{cases} & \begin{cases} (q-q')\cdot j_A = 0 \\ (q+q')\cdot j_A = -2mS \end{cases} \end{pmatrix}. \quad (29)$$

There are two conservation laws and two non-conservation laws in it applicable for the right antidiagonal currents. The "pseudoscalar" matrix presented as follows:

$$\begin{pmatrix} \begin{cases} (p+p')\cdot j_A = 0 \\ (p-p')\cdot j_A = -2mP \end{cases} & \begin{cases} (q+p')\cdot j_R = mP \\ (q-p')\cdot j_L = mP \end{cases} \\ \begin{cases} (p+q')\cdot j_R = -mP \\ (p-q')\cdot j_L = mP \end{cases} & \begin{cases} (q+q')\cdot j_V = 0 \\ (q-q')\cdot j_V = 2mP \end{cases} \end{pmatrix}. \quad (30)$$

It contains four non-conservation laws. As it can be seen from the structure of matrices obtained, the dual structure was related to the space-time structure - there are the vector and axial currents on the diagonal and on the antidiagonal - there are only the chiral ones. Now we need to determine whether there are the conditions, the implementation of which will transform the laws of non-conservation in the conservation laws. Such conditions are not difficult to find. They are contained in the diagonal equations of the two matrices. It is necessary to compare the upper diagonal equations of the "scalar" matrix to the lower diagonal equations of the "pseudoscalar" matrix. The result is the same in the reverse comparison.

Firstly, we compare the equations with the sums of impulses. They imply that the scalar $S$ is equal to zero if the sums of external and internal impulses happen to be proportional. The consequence of this will be conservation the of the right antidiagonal current. Nevertheless, we have to check whether the found condition correlates with the mirror symmetry. We will write down the condition for zero impulse components: $p_o + p'_o = \alpha(q_o + q'_o)$. To ensure the mirror symmetry, the factor $\alpha$ must take both positive and negative values and all the quantity of the impulses, corresponding with the condition, and must contain both positive and negative values of zero components of the external and internal impulses. Thus, it is impossible as the system (25) has a solution only with the positive energies. Therein, the condition can not be found, i.e. destroys the mirror symmetry, and hence, the system (25) can not conserve the right current.

We now compare the equations to the impacts. They imply that the pseudoscalar $P$ becomes zero if the external and internal impacts are proportional. The consequence of this is the emergence of the four conserved currents: two left antidiagonal and two diagonal. This means that the external equation begins to conserve the internal current and the internal equation conserves the external current, or what is the same, the left and right diagonal currents begin to conserve separately. Since the energy of the external impact as well as of the internal can have any signs, the found condition exposes mirror symmetry. Thus, in our study we have answered the question above – the dual bispinor generates four conventional conservation laws: two antidiagonal and two diagonal.

# 6. A dual electron

Now we consider the transition from a system of the spinor equations (25) to the corresponding system of the wave equations. Therefore, we use the rule (11). The first equation of the system(25) becomes the Dirac equation, which, in terms of the dual structure, naturally called the equation of *external electron wave*. We denote this wave as $\psi_p(x_p)$, where the coordinates of the external configuration space are denoted through $x_p^\mu$. The second equation becomes the equation of *internal electron wave*, which we denote by $\psi_q(x_q)$, where the coordinates of the internal configuration space are denoted through $x_q^\mu$. As a result of both transitions, the following system of the wave equations is to occur:

$$\begin{cases}(i\gamma_V \cdot \partial_p - m)\psi_p(x_p) = 0 \\ (i\gamma_A \cdot \partial_q + m)\psi_q(x_q) = 0\end{cases}. \quad (31)$$

The physical system provided by these wave equations is denoted as the *dual electron*. Thus, the last, described by two waves, kinematically and inextricably linked with each other, as generated by the same dual bispinor $W$. It is, therefore, natural to introduce the dual wave composed of inner and outer dual electron waves on the basis of the concept of the *dual isospinor* composed of two dual bispinors:

$$\begin{pmatrix}\psi_p(x_p) \\ \psi_q(x_q)\end{pmatrix}. \quad (32)$$

Now we state that the dual electron described by the dual wave and the Dirac wave is part of a dual wave. On comparing of (25) and (31), we see that the system (31) gives the wave expression of the dual symmetry, which is contained in the spinor form of the system (25), on the other hand, the system(25) represents a equation of the dual electron in the impulse representation.

# 7. Conserved currents of the dual electron

Let's consider the structure of the wave currents of the dual electron. Transition from the dual spinor currents to the dual wave currents is produced by the rule(11) and leads to the situation when the wave currents obtain explicit dual structure defined by the wave functions. This structure is naturally combined with the dual structure of the matrices (29) and(30), which, in turn, corresponds to the components of the dual isospinor. The diagonal wave currents describe transitions without changing the wave, i.e., correspond to the movements - external and internal, and the antidiagonal currents change the type of wave and correspond to the conversion processes. All the conclusions on the unconditioned and conditioned conservation laws of the spinor currents are transferred to a plane-wave currents unchanged. We write the remaining six conditionally conserved or resonance plane-wave currents in the form of the following matrix, using the notation $j_R^\mu = r^\mu, j_L^\mu = l^\mu$:

$$J_\mu = \begin{pmatrix} r_\mu^p + l_\mu^p & l_\mu^+ \\ l_\mu^- & r_\mu^q - l_\mu^q \end{pmatrix}. \quad (33)$$

Here the indices $p, q, +, -$, mark external, internal, raising and lowering currents respectively. From the form of this matrix it follows that the basis of the conserved currents structure of the dual electron is the algebra Li generators of the group $U(1)_R \times SU(2)_L$, acting in the dual isospin space (32).

# 8. Parity violation by a dual electron

Let us finally turn to the issue of mirror symmetry and its violation by the dual electron. In comparison to what we have – it clearly violates mirror symmetry matrix of the currents (33) - and then, taking in consideration the point where we started – an obvious mirror symmetric system of equations (25) - we see the paradoxical contradiction requiring its solution. The root of this contradiction contains in the matrices (29) and (30), so we leave matrix currents (33) for some time and introduce the following definitions. A reflection of three spatial coordinates will be called P- *reflection* and the symmetry about it called P-*symmetry*. A system of mathematical equations will be defined as *the theoretical* P-*symmetry*, if P-reflection goes into itself in this case. The physical phenomenon under observation called having *an experimental* P-*symmetry*, if the P-reflection turns into the same physical phenomenon observed.

Let us now consider the matrices (29) and (30), paying attention only to the conversion currents and considering the chiral and resonance condition: $P = 0$. As for them, it is stated that only left conversion currents conserved. While making P-reflection, we see that the upper conversion equations pass into the lower and lower - into the upper. As a result, after the P-reflection, we get the same system of equations, and the same left conserved current. Consequently, the matrices (29) and (30), and the system of equations (25) have a theoretical P-symmetry. As it is easy to see that the system with the left conserved currents is P-symmetric because under the P-reflection when the left current goes into the right, it loses its conservation and the right, going into the left, it acquires. Thus, P-reflection changes both chirality and conservation, and as a result, they are inextricably linked to each other – the left current is always conserved and the right - never. The essence of this paradox is that the only difference between the external and internal impulses generates an impact and, thus, the law of conservation so that the sum of impulses of the conservation law does not.

If we imagine that the observed electron is described by equations of the dual electron, then, having the theoretical P-symmetry, it will disrupt the experimental P-symmetry. At the same time we must imply a rule, as we have done before, that exists in nature and presupposes that only conserved currents can be observed. Then, the observed dual electron with the experimental P-symmetry and the P-reflection, the left observed current is expected to pass into the right one, and therefore, into the conserved current, which is impossible. The cause for differences of theoretical and experimental P-symmetries of the dual electron is considered to be the discharge of the observed currents from P-symmetric mathematical structure that occurs in violation of the P-symmetry.

Returning to the current matrix (33), we should say that it causes only the *appearance* of contradiction. This happens as a result of representation of only one part of a complete pattern given by matrix (30), and the one that indicates only the observed currents and, thereby, exhibits experimental violation of mirror symmetry. The right, non-conserved currents, providing theoretical mirror symmetry, can not be represented in the matrix(33), so this symmetry is hidden.

Thus, we found the dual electron consistent but paradoxical - having mirror symmetry *theory*, the electron breaks it *experimentally*.

# 9. Internal wave

Having solved the second of the equations (31), let's receive a plane internal wave of a dual electron in a view

$$\psi_{+q}(x_q) = e^{-iq \cdot x_q} W. \tag{34}$$

Let's find out physical sense of a wave vector of this wave. Its definition as internal impulse $q^\mu = ma^\mu$ gives nothing, as sense of a vector of polarisation $a^\mu$ in the existing theory is unknown. An exit in this situation is use of dual symmetry. We know the sense of a wave vector of an external wave is a vector of energy impulse of an electron $p_\mu$. To transfer this sense on an internal wave vector, it is necessary to make such definition of this concept which would be *dual invariant*. For this purpose we should proceed in this definition not from relativistic mechanics of our outside world, and from a structure of the wave. Let's make such definition: the wave vector of a wave is a vector of energy impulse of that particle which is described by a wave; energy is that part of a wave vector which remains in rest; an impulse is that part of a wave vector which in rest disappears.

Having applied this definition to an external wave, we will receive

$$p^\mu = (E, \mathbf{p}). \tag{35}$$

Let's apply it now to internal wave:

$$q^\mu = (p, \mathbf{E}). \tag{36}$$

Such is a consequence of dual symmetry: internal energy of a dual electron is a three-dimensional vector, and an internal impulse – a scalar. From a ratio (16)

$$p \cdot q = Ep - \mathbf{p} \cdot \mathbf{E} = 0, \tag{37}$$

from where we receive

$$p = \frac{\mathbf{p} \cdot \mathbf{E}}{E}. \tag{38}$$

The internal impulse is completely defined by external movement and internal energy. Internal movement is not independent from external: internal energy - is the external energy redistributed in the anisotropic image. At external rest there can't be internal movement.

## 10. Internal time and space

Let's find out now sense of internal configuration space, using the method applied above. Let's make such *wave definition* of time and space: time is that co-ordinate of configuration space which in a phase of the plane wave extending in this space, is increased by energy; space is that co-ordinate which is increased by an impulse.

Let's apply this definition to an external wave:

$$x_p^\mu = (t, \mathbf{r}). \tag{39}$$

Now we will apply it to an internal wave:

$$x_q^\mu = (r, \mathbf{t}). \tag{40}$$

Internal time of a dual electron appears three-dimensional, and internal space - one-dimensional. The dual plane wave describing a free dual electron, looks like:

$$\begin{pmatrix} \psi_{+p}(x_p) \\ \psi_{+q}(x_q) \end{pmatrix} = \begin{pmatrix} e^{-i(Et - \mathbf{p} \cdot \mathbf{r})} W \\ e^{-i(pr - \mathbf{E} \cdot \mathbf{t})} W \end{pmatrix}. \tag{41}$$

Accepted as it is represented, interpretation of internal time can be received if to start with its wave nature. The structure of a plane internal wave allows to assume that tridimentionality of internal time has potential sense, defining the possible directions of its current, actually it has one direction which is set by a vector of internal energy. Let's copy expression for a plane wave, having allocated in it weight of electron:

$$\begin{pmatrix} e^{-imv \cdot x_p} W \\ e^{-ima \cdot x_q} W \end{pmatrix} = \begin{pmatrix} e^{-im\tau_p} W \\ e^{-im\tau_q} W \end{pmatrix}. \tag{42}$$

From this it follows that plane waves of a dual electron represent two fluctuations with own frequency $m$ and own times $\tau_p$ and $\tau_q$. It is important that the role *of arrows of own times* for these fluctuations is played by speeds $v_\mu$ and $a_\mu$, being polarizations (21) of dual bispinor $W$. Therefore dual bispinor $W$ completely defines a structure of a plane dual wave and is for it *a bispinor of own times*.

Thus, we come to the general conclusion that in the theory of a dual electron the role of Minkowsky space time should be changed in comparison with the theory of a Dirac electron. This change as it is obvious, mentions only physical sense of co-ordinates in Minkowsky space, without concerning its mathematical structure. Apparently, it is simplest to make this modification, having entered concept *of Minkowsky variety*. It is necessary to understand the mathematical four-dimensional pseudo-Euclidean space which co-ordinates in itself have no physical sense as they are. The physical sense at co-ordinates of Minkowsky variety arises only when the bispinor wave extends in it and this sense is defined by this wave. For an external wave the Minkowsky variety is meaningful (39), for an internal wave − (40). Now we can eliminate an index in designation of configuration space and enter such definition of a plane dual wave on variety Minkovsky $x^\mu$:

$$\Psi_{+pl}(x) = \begin{pmatrix} \psi_{+p}(x) \\ \psi_{+q}(x) \end{pmatrix}. \tag{43}$$

It is necessary to mention that it is necessary to make reflexion of spatial axes for ensuring mirror symmetry according to their physical sense. Then, as well as it is necessary, the external wave will extend in vector space, and internal − in the pseudo-vector.

## 11 . Comparison to experience

In conclusion, we should say that we have almost reached the goal - the spatial structure of conserved currents of the dual electron coincides with the structure of the interacting currents of the observed electron, except for the weak neutral current, which is observed in the electron that is not completely axial. Therein, this difference is probably apparent. Let us compare, for example, the structures of the currents in Møller scattering of two dual electrons and two observed electrons described by the Standard Model. The dual electron current must have the left current including the electrodynamic vector current and axial current. Without going into the problem of the dual wave charges we can assume that it should effectively be neutral. In this case, the neutral non-vector current dual electron will consist of the axial and left currents and its structure can be quite consistent with that of the observed electron at the value of the Weinberg angle $\sin^2 \theta_W < 1/4$. Moreover, the presence of the left neutral current may explain as to why the Weinberg angle is not equal to $30^\circ$.

## 12 . Discussion

In the real work the origin of remaining currents in the Dirac theory is studied. In work it is shown that if in the Dirac theory to start with spinor geometry and its simmetry, other structure of conserved currents, than in the picture of a Dirac electron based on the wave equation turns out. These differences are the following:

α) six currents remain;

β) parity is broken;

γ) the structure of currents is close to observed at a physical electron.

The consideration carried out in work represents immanent modification of the Dirac theory, i.e. such one, which, without changing the initial equations of the theory, gives them new senses, developing existing in the theory of symmetry is so far, as far as it is possible. Work has heuristic character and its results represent only the first contours of a new *picture of a dual electron* which in the complete development should include a picture of a Dirac electron and, thereby, replace it.

The similarity of spatial structures of currents of a dual electron and electron of Standard model is combined with a dissimilarity of their physical senses. The unique possibility to pull them together as it seems now is the assumption of existence at of dual structure in neutrino. As mostly remaining currents of a dual electron are conditional, transition to a co-operating dual electron represents, apparently, yet not contemplated problem.

## 13 . Conclusion

We conclude the article commenting on the comparison of the dual electron and the electron of the Standard model. The dual electron is a mathematical construct, generated by the Dirac algebra and existing independently of experience. The electron of the Standard model is essentially taken from experience. If their proximity would not be accidental, it would mean that the world is ruled by algebra.